# Enhancing Control Performance through ESN-Based Model Compensation in MPC for Dynamic Systems

1st Shuai Niu
*School of Mechatronical Engineering and Automation，Shanghai University*
Shanghai, China
shuainiusn@163.com

2nd Qing Sun
*School of Mechatronical Engineering and Automation，Shanghai University*
Shanghai, China
qingsun@shu.edu.cn

3rd Minrui Fei
*School of Mechatronical Engineering and Automation，Shanghai University*
Shanghai, China
mrfei@staff.shu.edu.cn

4th Xuqian Ju
*School of Mechatronical Engineering and Automation，Shanghai University*
Shanghai, China
ju_xuqian@163.com

* Corresponding author: qingsun@shu.edu.cn

*Abstract*—Deriving precise system dynamic models through traditional numerical methods is often a challenging endeavor. The performance of Model Predictive Control (MPC) is heavily contingent on the accuracy of the system dynamic model. Consequently, this study employs Echo State Networks (ESN) to acquire knowledge of the unmodeled dynamic characteristics inherent in the system. This information is then integrated with the nominal model, functioning as a form of model compensation. The present paper introduces a control framework that combines ESN with MPC. By perpetually assimilating the disparities between the nominal and real models, control performance experiences augmentation. In a demonstrative example, a second- order dynamic system is subjected to simulation. The outcomes conclusively evince that ESN-based MPC adeptly assimilates unmodeled dynamic attributes, thereby elevating the system's control proficiency.

## I. INTRODUCTION

This As contemporary industrial systems grow increasingly com- plex, the demand for efficient, stable, and reliable control strategies has surged. MPC, as an advanced control approach, has demonstrated significant achievements across various applications. By optimizing the system's behavior over a future time horizon, MPC empowers the controller to make optimal control decisions at each time instant [1]. However, practical applications of MPC face challenges related to model accuracy, especially in complex, nonlinear systems, where accumulated model errors can lead to degraded control performance or even instability [2]. To ad- dress this issue, model compensation has emerged as a critical research direction. Model compensation introduces real-time model corrections into the MPC control loop, rectifying model inaccuracies and thereby enhancing control performance and robustness [3].

With the advancements in artificial intelligence and deep learning technologies, control methods based on neural networks have garnered increasing attention. Among these, the ESN, characterized by its unique recur- rent neural network architecture, has demonstrated outstanding performance in time-series data modeling and dynamic system control. It enables the recovery of system model components overlooked due to linearized modeling, employing a data- driven approach [4].

The ESN was originally proposed by Jaeger et al. and has achieved significant success in fields such as time series prediction and system identification [5] [6] [7]. In prior research on ESN, many researchers treated ESN as an alternative model and proposed model-free control methods. For instance, in [8], the initial system identification process was omitted, and an end-to-end control method was constructed directly using the system's input and output. In [9], a parallel computing scheme based on ESN was proposed for model-free prediction of large-scale spatiotemporal chaotic systems. Such methods offer unique advantages for abstract and unmodeled systems. However, considering prior information about the system model can contribute to a better approximation of the real model. In [10], the study considered the model of mobile robots in the real world, where the prior part of the dynamics was derived using conventional numerical methods, and the unmodeled part was trained using Gaussian processes for regression. The Echo State Network, as a regression model, can also fit the model compensatory part from training data. In [11], a novel ESN network structure was proposed for controlling uncertain SISO nonlinear systems. Additionally, it's crucial to note that for ESN, hyperparameter optimization significantly influences the network structure, consequently impacting the ultimate predictive performance [12].

This paper begins by presenting the problem model in section II, along with an exposition of the foundational principles underpinning MPC and ESN. Section III elaborates on ESN-based MPC, providing a flowchart delineating the control approach. In section IV, a straightforward second- order system is devised, accompanied by the specification of pertinent parameters. Section V entails a comprehensive examination of simulation outcomes, unequivocally affirming the efficacy of the proposed algorithm.

## II. BACKGROUND

### A. Problem Formulation

In this paper, we consider a discrete dynamic model:
$$x(k+1) = f_{true}(x(k), u(k))$$

where $f_{true}$ denotes the true model, $x(k) \in \mathbb{R}^{n_x}$ and $u(k) \in \mathbb{R}^{n_u}$ are the system state and input, respectively. assume that the model can be represented by the following equation:
$$x(k+1) = f_{nom}(x(k), u(k)) + B_n(d_{esn}(x(k), u(k)))$$

where $f_{true}$ denotes the nominal model, the nominal model is typically obtained by locally linearizing around a specific state point, such as the Runge-Kutta method. The matrix $B_n \in \mathbb{R}^{n_x \times n}$ determines the state's subspace influenced by the residual dynamics, $d_{esn}$ denotes the unmodeled feature terms regarding the state and input.

### B. Standard MPC Formulation

MPC is fundamentally based on predicting the future dynamic behavior of a system within a finite prediction horizon (often referred to as N steps), and selecting control inputs to minimize a performance metric. By solving an optimization problem, the first value of the optimal control input sequence is applied to the system. The control inputs in the prediction horizon $U = [u(0), u(1), \dots, u(N-1)]$ and predicted state trajectories $X = [x(0), x(1), \dots, x(N)]$ are all constrained in the respective sets $\mathcal{X}$ and $\mathcal{U}$ with the terminal set $\mathbb{X}_f$. This can be represented in the following form.

$$\min_{X,U} V_N = \sum_{k=0}^{N-1} l(x(k), u(k)) + V_f(x(N))$$

**subject to**: $x(0), u(k-1)$ given:
$$x(k+1) = f_{nom}(x(k), u(k)),$$
$$x(k) \in \mathcal{X}, \ k = 0, 1, \dots, N-1$$
$$u(k) \in \mathcal{U}, \ k = 0, 1, \dots, N-1$$
$$x(N) \in \mathbb{X}_f$$

where the cost function $l(x(k), u(k))$ and the final cost $V_f(x(N))$ are defined according to the considered control problem.

### C. Echo State Network

ESN is a neural network architecture specifically designed for modeling time-series data and dynamic system control. It incorporates internal feedback loops to retain and utilize historical state information, thereby enhancing the network's capability to model dynamic systems. ESN has demonstrated outstanding performance in various tasks related to time-series prediction and control, particularly in systems exhibiting complex dynamic behaviors. The primary focus lies in training the weight matrix $W_{out}$ that connects the reservoir to the output in ESN. The mapping of inputs to the reservoir ($W_{in}$) and the weights connecting internal nodes of the reservoir ($W_{res}$) are pre-defined and remain fixed. Therefore, much like other neural networks such as Recurrent Neural Network (RNN), Convolutional Neural Network (CNN), Long Short-Term Memory (LSTM), the selection of hyperparameters is crucial. It consists of three layers: input layer, hidden state layer, and output layer. The state update of reservoir nodes in the hidden layer is described by the following equation:

$$s_i(t+1) = f\left(\sum_{j=1}^{N} W_{res}(i,j) s_j(t) + \sum_{k=1}^{M} W_{in}(i,k) x_k(t)\right)$$

where reservoir contains $N$ nodes, and the state of each node is represented by $s_i(t)$, where $i = 1, 2, \dots, N$, $W_{res}(i,j)$ denotes the weight connecting the $i$-th reservoir node to the $j$-th reservoir node, $W_{in}(i,k)$ denotes the weight linking the $k$ th input layer node to the $i$-th reservoir node, and $W_{out}(i,j)$ denotes the weight connecting the $i$-th output node to the $j$-th reservoir node, $f$ is the activation function, typically chosen as the sigmoid function or hyperbolic tangent function.

The computation of the output layer is described by the following equation:

$$y_i(t) = \sum_{j=1}^{N} W_{out}(i,j) x_j(t)$$

where $W_{out}(i,j)$ is the weight connecting the $i$-th output node to the $j$-th reservoir node.

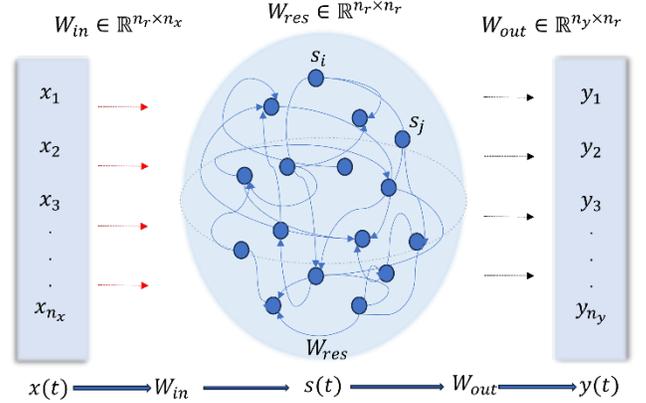

Fig. 1. At each time instant $t$, the reservoir of the ESN consists of $N_r$ nodes. The dimension of the input $x(t)$ depends on $n_x$, and the dimension of the output $y(t)$ depends on $n_y$.

The fundamental concept of Echo State Networks involves the multiplication of input data by a fixed random weight matrix. This leads to dynamic evolution within a high-dimensional state space, enabling the modeling of complex nonlinear problems. Fine-tuning network parameters, such as network size, spectral radius, and sparsity, can enhance the network's performance in addressing specific problems. In the context of the control framework presented in this paper, the optimization of network hyperparameters is not the primary

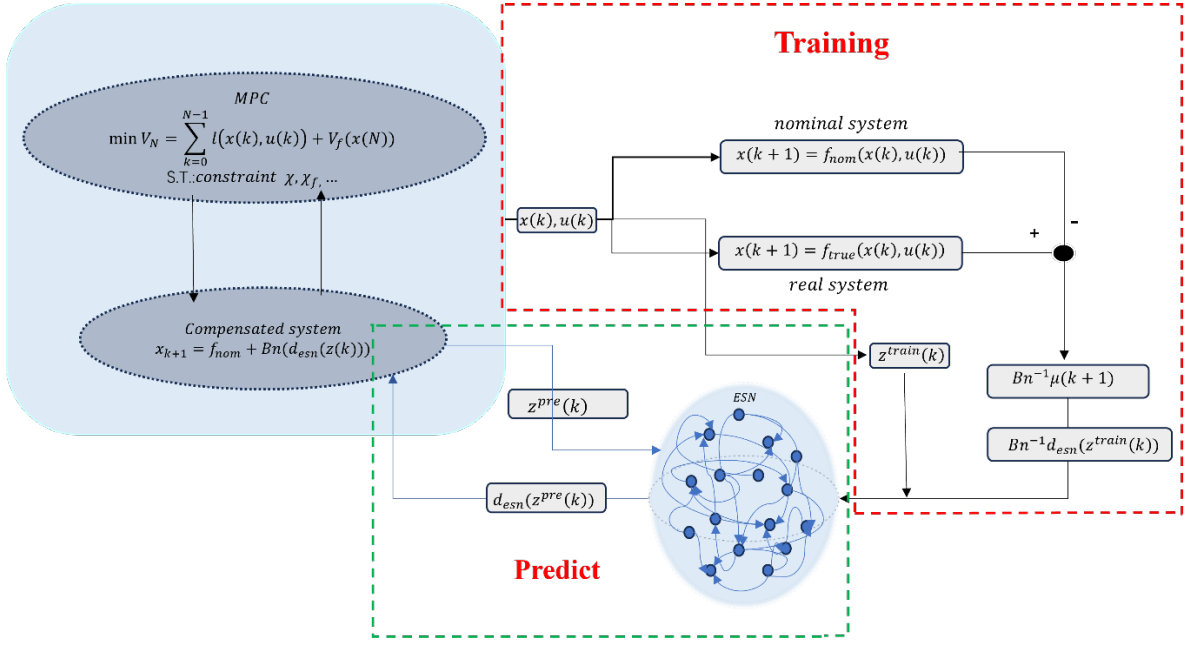

Fig. 2. Diagram illustrating the procedural flow of the MPC framework utilizing ESN. Initially, the error compensation model serves as the system dynamics constraint for MPC (utilizing the nominal model during the data collection phase). Subsequently, the optimally computed control input is administered to both the true system and the nominal system. The disparity between the newly evolved state values of the two systems is computed and subjected to weighting before integration into the ESN. This facilitates the ESN in learning the unmodeled error components.

focus. Instead, an empirical approach is employed to select suitable hyperparameters

## III. ESN-BASED MPC

In this section, we present the combined framework of ESN and MPC. After linearizing a nonlinear system, unmodeled dynamics can potentially degrade the effectiveness of MPC, as the predictions may not accurately represent the true system dynamics. To address this, the ESN can be employed to learn the unmodeled dynamics, thereby enhancing the predictive capabilities of MPC. The focus lies on the learning of $d_{esn}$ in (1). Equation (2) can be transformed into the following form:

$$B_n(d_{esn}((x(k),u(k)))) = x(k+1) - f_{nom}(x(k),u(k))$$

It is typically unnecessary for the state vector $x$ to have a large dimension. A common practice is to select the most relevant state variables by setting the weight matrix $B_n \in \mathbb{R}^{n_x \times n}$ and $B_z \in \mathbb{R}^{n_z \times (n_x + n_u)}$. This helps reduce the computational burden generated during the training process, thus accelerating the algorithm's execution. Further simplification can be achieved as follows:

$$d_{esn}(z(k)) = B_n^{-1}(\mu(k+1))$$

where $z(k) = [B_z(x_{\text{true}}(k), u(k))], \mu(k+1) = x_{\text{true}}(k+1) - x_{\text{nom}}(k+1)$ denotes the difference at time $k+1$ between the true state $x_{\text{true}}(k+1)$ of the system obtained through measurement and the nominal state $x_{\text{nom}}(k+1)$ calculated by the nominal model. In this method, the discrepancy in state variables between the nominal system and the true system is employed as the target label $Y$ for training the ESN. We begin by employing the nominal model as the constraint model for MPC. Following the establishment of fixed hyperparameters for ESN, we quantify the disparity between the actual system and the nominal system using the optimal input determined by MPC. This discrepancy $(z^{train}, B_n^{-1}d_{esn}(z^{train}(k)))$ is then incorporated into the training dataset. After the completion of training, we discard the error model and, in its place, utilize the state $z^{pre}(k)$ quantities computed by the compensation model to forecast the remaining dynamics $d_{esn}(z^{pre}(k))$ of the system.

It is assumed that observations of the true system are impeccably precise, devoid of any observational noise. The state values and input values obtained under the error compensation system serve as the training data $X$. This approach enables the development of a dynamic model depicting the evolution of error terms under different state and input conditions. The specific algorithmic framework is illustrated in Figure 2.

## IV. EXAMPLES

In this section, a simulation example of a second-order spring-damping system is provided. It compares the performance of MPC under different models with the predictive performance of ESN. The schematic diagram of the spring-damping system is shown in Figure 3.

The displacement equation describing the change in position over time and the velocity equation describing the change in velocity over time for the spring-damping system are given as follows:

$$m\ddot{x}(t) + b\dot{x}(t) + kx(t) = F(t)$$
$$\dot{x}(t) = v(t)$$

where $m$ is the mass, $b$ is the damping coefficient, $k$ is the spring constant, $x(t) =$ is the position $s$ as a function of time, $\dot{x}(t)$ is the derivative of velocity $v$ with respect to time $(dx/dt)$,

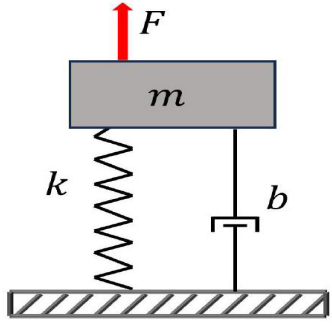

Fig. 3. In this system, there is a mass $m$ object, with a spring constant $k$ and a damping coefficient $b$. The motion of the object can be altered by applying an external force $F$

$\ddot{x}(t)$ is the derivative of acceleration $a$ with respect to time $(d^2x/dt^2)$, $F(t)$ is the external force. $v(t)$ is the velocity as a function of time.

The discrete state-space equations of a second-order spring-damping system can be obtained by discretizing the continuous-time state-space equations. Assuming a sampling time interval of $dt$, the discrete state-space equations are as follows:

$$x(k+1) = Ax(k) + Bu(k)$$

$$A = \begin{bmatrix} 1 & dt \\ 1 - \frac{k}{m}dt & 1 - \frac{b}{m}dt \end{bmatrix}, B = \begin{bmatrix} 0 \\ \frac{dt}{m} \end{bmatrix}$$

where the states $[s, v]$ and $u = F$, the considered springdamping system involves applying a force $f$ to the mass block in order to modify the velocity and position of the spring. For this simplified second-order model, let's assume $Bz^T = [1,1]$, which implies taking into account the influence of all state variables on the error term. The example is constructed such that it is possible to display the resulting ESN in 2D output. The true model which is actually defined by:

$$x_{\text{true}}(k+1) = f_{\text{nom}}(x_k, u_k) + B_n(d_{esn}(z(k)))$$

In our example the parameters of the spring-damping system were set to m = 1.0, k = 10, b = 0.5. The implemented MPC has a prediction horizon N = 20 and the control formulation aims to stabilize the mass at $s = 0$ with $v = 0$. The loss function $l(x_k, u_k)$ can be expressed as:

$$l(x, u) = (x - r)^T Q(x - r) + Ru^2$$

where the reference signal $r$ is set to $r = [0,0]$. The matrices $Q = [1, 0.1]$ and $R = [0.1]$ serve as weighting factors, penalizing the deviation between a subset of states $x$ and the reference signal $r$, as well as the magnitude of the input force $F$.

ESN is employed to learn the unmodeled perturbation component. The preset hyperparameters are: reservoir size $N^r = 1500$, regularization coefficient $\beta = 1 \times 10^{-4}$, sparsity $= 3$, spectral radius $\rho = 1$, and leaky rate $\gamma = 0.4$, washout $\lambda = 30$. In practical simulation, it can be observed that this is an underdamped system.

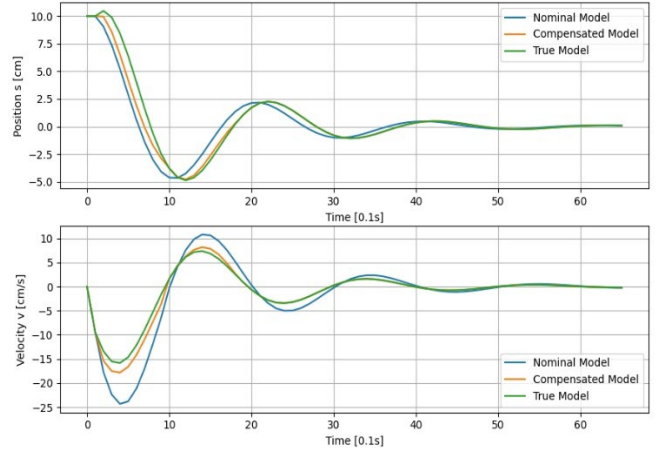

Fig. 4. The simulation results demonstrate that the compensated model is closer to the true model. Here, due to the selection of a second-order underdamped spring-damping system, it converges to the target point after a brief oscillation. In fact, the nominal model also performs well in this scenario. However, this simulation is conducted to substantiate the effectiveness of model error compensation.

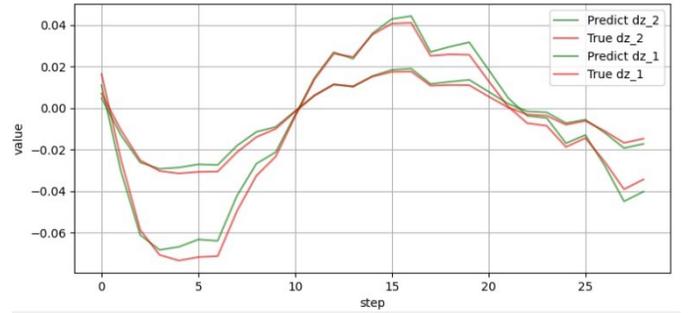

Fig. 5. In this simulation, $\lambda = 30$ was chosen. A total of 70 data points were used for training, and the results include predictions for 30 steps. Since the predictions are two-dimensional, it can be observed in the graph that the two prediction curves provided by ESN closely match the true values under the same conditions.

## V. RESULTS

The simulation time is set to 10 seconds with a time step of 0.1 seconds. Therefore, a total of 100 steps of simulation results were obtained. In this section, simulation results for an example of this spring-damping system are discussed. The simulation was conducted in a Python 3.8 environment. The initial state of the system was set as $x = [10,0]$ and the target point was defined as $r = [0,0]$. The simulation was executed for a duration of 10 seconds, employing a time step of 0.1 seconds. Consequently, a total of 100 simulation steps were recorded.

Initially, the standard MPC was implemented using the nominal model to collect training data. During this phase, the ESN was not subjected to training. Once a sufficient amount of training data was accumulated, the data was subsequently input offline into the ESN. Meanwhile, during the training of the ESN, the washout $\lambda$ was set to 10 to facilitate the initialization of the reservoir nodes. After the completion of the entire training process, the weight matrix of the ESN was obtained.

In the subsequent run, the model error compensation term $B_n(d_{esn}(.))$ obtained from training the ESN was incorporated into the nominal model. MPC was then executed again. This approach was pursued to both observe the system's performance under the influence of the error compensation model, and concurrently, continuously gather new data for online updating of the reservoir nodes within the ESN network. The simulation results for the spring-damping system are shown in Figure 4. Similarly, Figure 5 illustrates the predictive results of the ESN for the training values after undergoing a warm-up period with $\lambda = 30$.

## VI. CONCLUSIONS

This paper introduces a control framework that integrates Echo State Networks (ESN) with Model Predictive Control (MPC). Given the substantial challenge in deriving precise system dynamics models through traditional numerical methods, ESN is employed to learn unmodeled system errors. Nonetheless, it is worth noting that the nominal model often captures a significant portion of the system's characteristics. As a result, a combination of the nominal model and error compensation is explored. The effectiveness of this approach is demonstrated through a simple second-order system, showcasing significant enhancements in the performance of the final controller. However, it is important to acknowledge that this paper does not delve into the optimization of hyperparameters for ESN, which could potentially lead to further improvements in predictive accuracy. Moreover, it is essential to acknowledge that the scope of this paper is limited to relatively straightforward second-order systems. For more complex real-world models, additional refinement of this control methodology is warranted.